# Ultrafast X-ray induced damage and nonthermal melting in cadmium sulfide

Nikita Medvedev[1,2,*], Aldo Artímez Peña[1,3]

1) Institute of Physics, Czech Academy of Sciences, Na Slovance 1999/2, 182 00 Prague 8, Czech Republic

2) Institute of Plasma Physics, Czech Academy of Sciences, Za Slovankou 3, 182 00 Prague 8, Czech Republic

3) Higher Institute of Technologies and Applied Sciences, University of Havana, Cuba

## Abstract

Cadmium sulfide is a valuable material for solar cells, photovoltaic, and radiation detectors. It is thus important to evaluate the material damage mechanisms and damage threshold in response to irradiation. Here, we simulate the ultrafast XUV/X-ray irradiation of CdS with the combined model, XTANT-3. It accounts for nonequilibrium electronic and atomic dynamics, nonadiabatic coupling between the two systems, nonthermal melting and bond breaking due to electronic excitation. We find that the two phases of CdS, zinc blende and wurtzite, demonstrate very close damage threshold dose of ~0.4-0.5 eV/atom. The damage is mainly thermal, whereas with increase of the dose, nonthermal effects begin to dominate leading to nonthermal melting. The transient disordered state is a high-density liquid, which may be semiconducting or metallic depending on the dose. Later recrystallization may recover the material back to the crystalline phase, or at high doses create an amorphous phase with variable bandgap. The revealed effects may potentially allow for controllable tuning of the band gap *via* laser irradiation of CdS.

## I. Introduction

Cadmium sulfide is a wide band gap semiconductor that found applications in solar cell photovoltaic technologies [1]. Various optoelectronic devices are based on CdS, including light-emitting diodes [2]. Due to the presence of a relatively heavy element, cadmium, it is also used in radiation detectors [2].

For practical purposes, controlled irradiation is often used for material processing and manufacturing of materials with desired properties [3–6]. Usage of CdS under high radiation doses poses a question of its stability, damage threshold, and damage mechanisms involved. Previous research on X-ray irradiation of CdS was performed in low flux (low dose-rate) regime, in which point defects were produced by individual photons [7,8]. However, high dose-rate irradiation may induce different kinetics and produce different damage [9].

---

[*] Corresponding author: nikita.medvedev@fzu.cz

Fundamentally, upon laser irradiation, the following processes take place in matter [10,11]: first, electrons absorb photons and get promoted to higher-energy states, such as excitation from the valence to the conduction band in band-gap materials. This process takes place during the laser pulse. If the photon energy is sufficiently high, e.g. for extreme ultraviolet (XUV) or X-ray lasers, the photons are mainly absorbed by core atomic shells, exciting bound electrons to free states and leaving core shell holes behind. The core holes decay via Auger (or radiative, for deep shells of heavy elements) process, which typically takes place at femtosecond timescales [12,13]. The excited electrons then scatter with matter exciting new electrons (impact ionization), with the collective electron modes (plasmons), and with atoms and their collective modes (phonons) transferring energy to the atomic lattice [11,13]. Electronic kinetics leads to equilibration of the electronic ensemble, establishing their equilibrium Fermi-Dirac distribution, typically also at femtosecond timescales.

Electrons transfer energy to the atoms via two main mechanisms: at picosecond timescales, nonadiabatic electron-ion (electron-phonon) coupling, and adiabatic modification of the interatomic potential which at high radiation doses may lead to nonthermal melting or bond breaking and ultrafast atomic disorder even without significant atomic heating [14,15]. At doses higher than the nonthermal melting threshold, the same process of electron-induces changes in the interatomic potential may lead to atomic acceleration and further ultrafast heating of atoms at sub-picosecond scales [16]. The atomic heating and bond breaking may induce phase transitions, forming new states of matter – different solid or liquid states, or even transient unusual states outside of the equilibrium phase diagram [17,18].

To study CdS response to ultrafast radiation, we will apply a state-of-the-art simulation combining a few models into one hybrid approach implemented in the code XTANT-3, which allows us to simulate all the abovementioned effects [19,20]. We will study the two most common phases of this semiconductor: zinc blende and wurtzite. The main aim of this work is to investigate the processes induced by ultrafast intensive XUV/X-ray irradiation, find the material damage mechanisms and damage threshold, and assess the material states produced in phase transitions. We will demonstrate that the irradiation may allow for controllable tuning of the material band gap, varying the states between semiconducting and metallic liquid/disordered states.

## II. Model

Damage kinetics in CdS induced by ultrafast X-ray or XUV radiation is simulated with the hybrid (multiscale) code XTANT-3 [19]. The code concurrently links multiple models for description of various processes [21]. The full detail of the code can be found, e.g., in Ref. [22]; here, we provide a brief description of the physical effects included in the model and the methodology of their numerical description.

The X-ray/XUV photon absorption, instigated electron cascades, and Auger decays of core holes, if any, are modelled with the transport Monte-Carlo (MC) method of event-by-event (analog) simulation [20,21,23]. Photoabsorption cross sections, Auger decay times, and ionization potentials of core shells are extracted from EPICS2023 database [24]. The kinetics of electrons with the kinetic energy above the chosen cutoff of 10 eV, counted from the bottom of the conduction band, are modelled with the screened Rutherford cross-section with the modified Molier screening parameter for elastic scattering [23]; the Ritchie-Howie complex-dielectric-function (CDF) formalism is used for the inelastic scattering (impact ionization of core holes and valence band, and scattering on plasmons) [25]. The parameters of the CDF for CdS are obtained within the single-pole approximation [26]. The MC simulation is averaged over 10,000 iterations to obtain reliable statistics [21,27].

Electrons with energies below the cutoff, populating the evolving valence and conduction bands, are traced with the Boltzmann collision integrals (BCI). In the present work, we consider them to adhere to the Fermi-Dirac distribution (instantaneous electron thermalization approximation in electron-electron scattering) [28]. The matrix element for the nonadiabatic energy exchange between these electrons and atoms (electron-phonon coupling) is extracted from the transient tight binding (TB) Hamiltonian [29].

The transferable tight binding method is used for the evaluation of the transient electronic orbitals (energy levels, band structure), as well as for calculation of the interatomic forces [30]. The transient Hamiltonian, dependent on the position of all the atoms in the simulation box, is diagonalized at each time step of the simulation, allowing us to trace evolution of the electronic structure and the atomic potential energy surface in response to electronic excitation. For CdS, we apply the DFTB parameterization from Ref. [31], which uses $sp^3d^5$ basis set for the linear combination of the atomic orbitals. Additionally, to trace high-temperature states of excited material, short-range repulsive potentials for Cd-Cd and Cd-S interactions are added based on the ZBL-potential, similar to Ref. [32].

Atomic motion is traced with the classical Molecular Dynamics (MD). The interatomic forces are calculated from the TB Hamiltonian and the transient electron distribution function (fractional electronic populations traced with BCI method above). This way, the model accounts for changes in the interatomic potential that rise from alterations in the electronic distribution due to X-ray-pulse excitation and high-energy electron scattering. Thus, the model can describe nonthermal melting: a phase transition induced by the modification of the atomic bonds due to electronic excitation, without the atomic heating [15,33]. The energy transferred in the nonadiabatic (electron-phonon) coupling, evaluated with the BCI, is delivered to the atoms *via* the velocity scaling algorithm at each timescale of the simulation [29]. Martyna-Tuckerman $4^{th}$ order algorithm is used for propagation of the atomic trajectories with the timestep of 0.5 fs [34]. We use 200 atoms in the simulation box with periodic boundary conditions for modelling the wurtzite structure of CdS, while 216 atoms are used for the zinc blende structure, which is sufficient for a reliable simulation [20]. The atomic coordinates in the unit cells of both structures are taken from Ref. [35]. The simulation begins a few hundred fs before the pulse arrival to allow for the atomic thermalization and is typically run up to 5 ps time after irradiation (with gaussian laser pulse centred at 0 fs) [36]. Atomic snapshots are visualized with help of OVITO [37].

XTANT-3 has been previously validated against experiments on damage kinetics in various irradiated materials, showing a reasonable agreement with experiments (see, e.g., Refs. [18,20,33,38]).

## III.    Results

### 1. Zinc Blende structure

We modelled irradiation of CdS with various doses to find the threshold of onsetting a phase transition. The material disorders at the dose of ~0.4-0.5 eV/atom, see atomic snapshots showing the material response to below and above the threshold dose in Figure 1. At the shown dose of 0.6 eV/atom, by the time of 500-1000 fs, the atomic lattice loses stability and turns into disordered liquid-like state.

In contrast, Born-Oppenheimer simulation, which excludes the electron-phonon coupling, and thus nonadiabatic heating of atomic system [39]; show that the damage onsets at the higher doses of ~0.8 eV/atom. The difference suggests that this phase transition is mainly of the thermal nature, induced by the atomic heating via electron-phonon coupling.

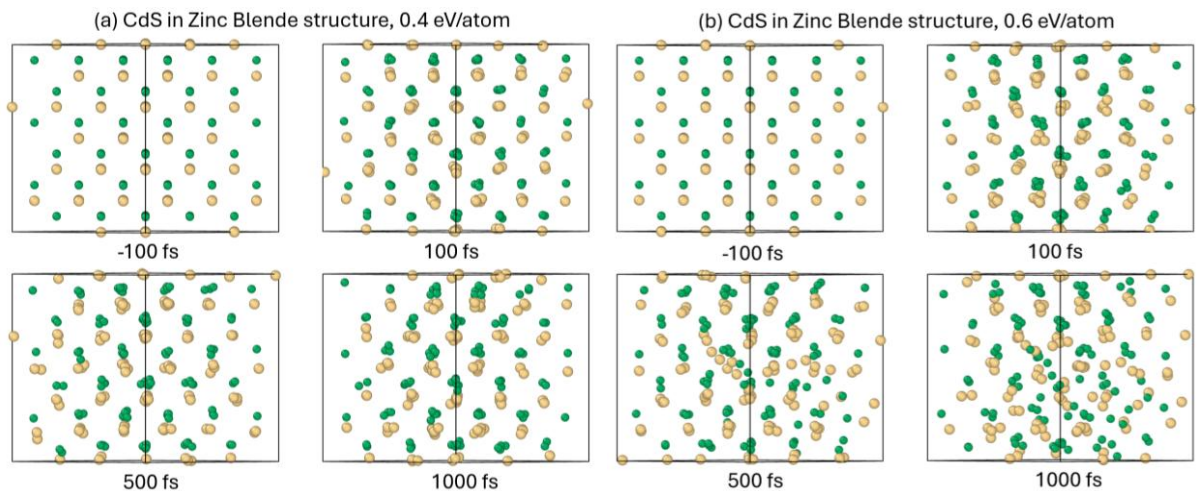

Figure 1. Atomic snapshots of CdS in *Zinc Blende* structure irradiated with (a) 0.4 eV/atom, and (b) 0.6 eV/atom absorbed dose, 10 fs FWHM, 30 eV photon energy pulse. Large yellow balls are Cd, small green balls are S.

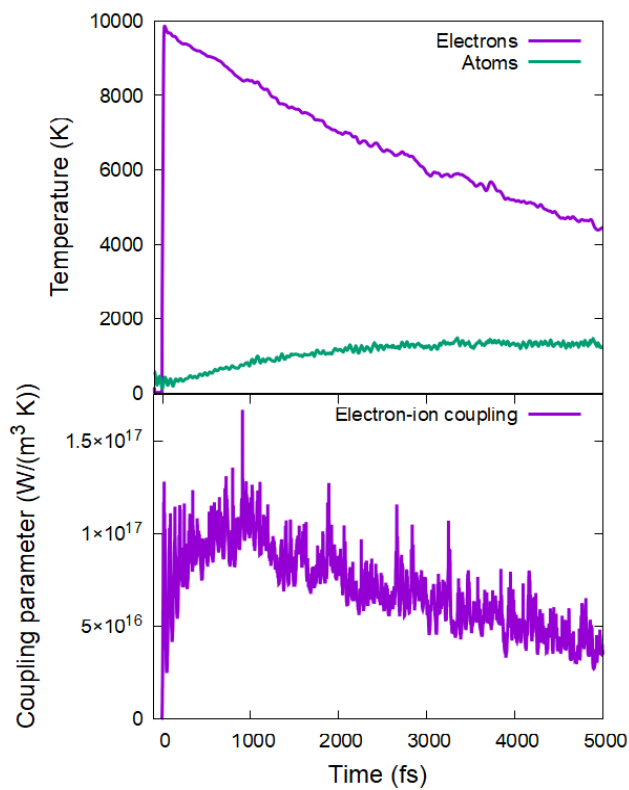

Figure 2. Electronic and atomic temperatures (top panel) and electron-ion coupling parameter (bottom panel) in Zinc Blende CdS irradiated with 0.5 eV/atom.

The atomic heating via electron-phonon coupling in non-BO simulations is relatively slow, see the example of the equilibration of the electronic and atomic temperatures in Figure 2. The coupling parameter reaches its peak of ~$10^{17}$ W/(m$^3$K) at the time of ~1 ps after the pulse when the electronic temperature is still relatively high, and the atomic temperature is also close to its maximum [40]. Afterwards, the coupling parameter decreases with decrease of the electronic temperature to low values [41]. The atomic temperature increases further only a little, since the energy transiently stored in the electronic system is small due to low electronic heat capacity [41].

In response to irradiation, the band gap of CdS shrinks with increase of the dose. As is typical for ionic materials, the threshold dose for atomic disorder is lower than the threshold dose of the complete band gap collapse [42]. Figure 3 shows that at the melting threshold dose, the band gap transiently contracts to ~1 eV; for a complete band gap collapse, longer timescales (~3 ps), or higher doses of ~1 eV/atom are required. Thus, XTANT-3 calculations predict that the irradiated CdS may transiently form two disordered (melted) states: semiconducting in the dose window from ~0.5 eV/atom to ~1 eV/atom, and metallic for the doses above ~1 eV/atom. Moreover, the results suggest that varying the irradiation dose, one may controllably tune the band gap, which may find its use for practical applications in designing CdS-based solar cells and electronics.

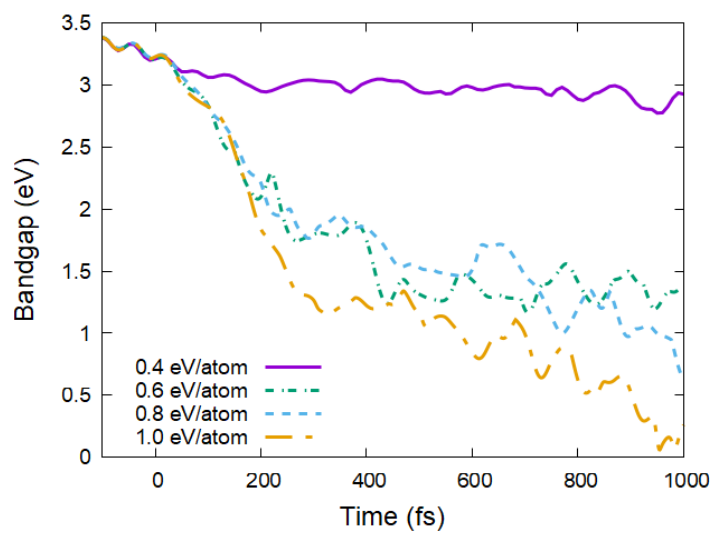

*Figure 3. Band gap of Zinc Blende CdS irradiated with various doses.*

The band gap collapse at high doses takes place mainly *via* lowering of the conduction band, whereas the valence band symmetrically widens, see Figure 4. That means that the conduction band electrons merging with the valence band holes due to energy levels shift lose their energy to nonthermal acceleration of the atoms [16].

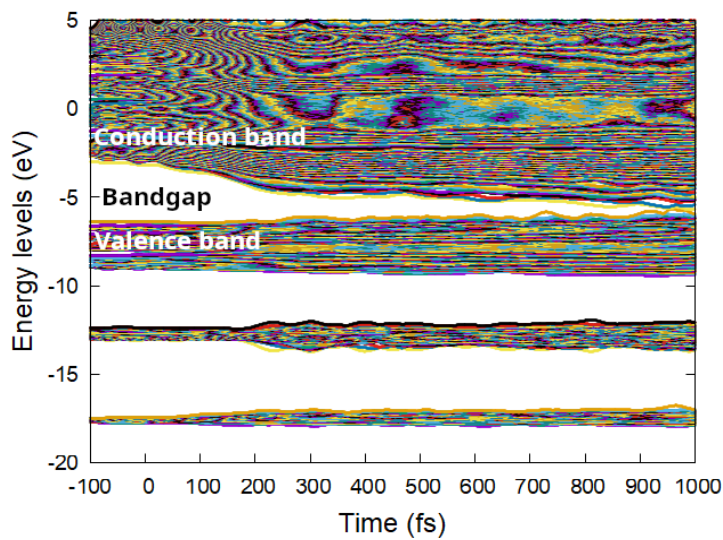

*Figure 4. Electronic energy levels (molecular orbitals, band structure) in Zinc Blende CdS irradiated with 1.0 eV/atom dose.*

It is also interesting to note that phase transition to the disordered state is accompanied by the pressure turning negative (stress), see Figure 5. This indicates that the material is attempting to contract, as confirmed by the NPH-simulation (not shown). Correspondingly, the density of the liquid CdS is higher than that of the crystalline Zinc Blende state. Similar formation of a high-density liquid state after irradiation was reported in silicon [39,43].

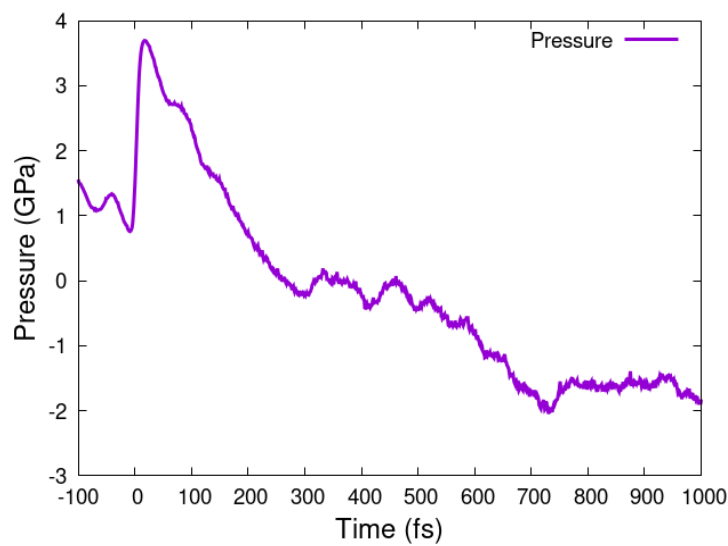

*Figure 5. Pressure in Zinc Blende CdS irradiated with 1 eV/atom dose.*

## 2. Wurtzite structure

CdS in Wurtzite structure also damages at the irradiated dose of ~0.4-0.5 eV/atom. The atomic snapshots in Figure 6 show that the material disorders at doses above this threshold. The processes taking place in Wurtzite are the same as in Zinc Blende structure of CdS – the disorder is thermal, induced by the electron-ion coupling heating up the atomic system at the timescale of 2-3 ps.

It is also accompanied by the band gap shrinkage and eventual complete closure at long timescales. The ultrafast complete collapse of the band gap (within 1 ps) takes place at the dose of ~1.1 eV/atom, see Figure 7. This suggests that Wurtzite CdS also has a range of doses where the irradiated material is, at least transiently, a semiconducting liquid (~0.5-1 eV/atom), while at higher doses it turns into a metallic liquid. The liquid CdS also has a higher density than the crystalline Wurtzite phase, as seen by the fact that the pressure turns negative at above-threshold irradiation doses, see Figure 8.

Thus, we conclude that both material phases behave similarly under ultrafast XUV/X-ray irradiation.

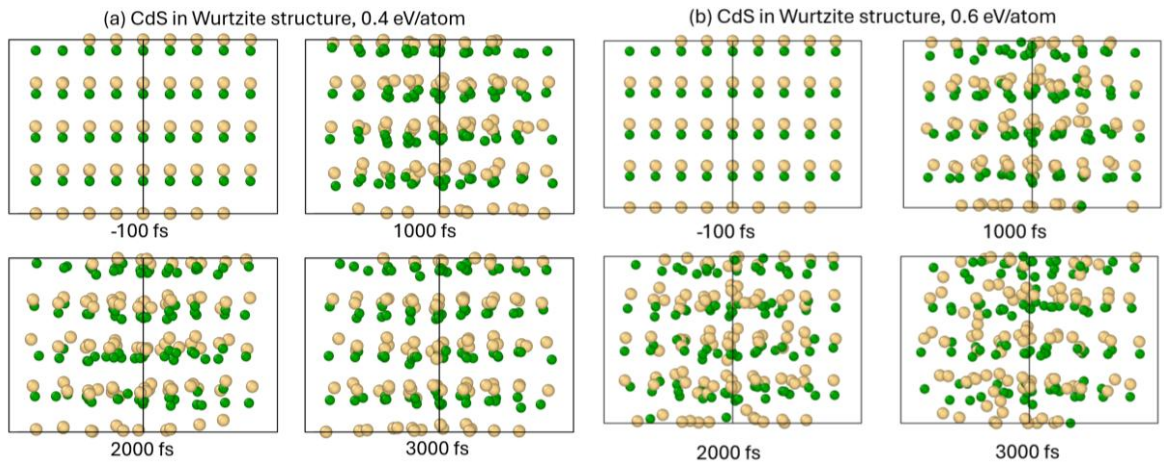

*Figure 6. Atomic snapshots of CdS in Wurtzite structure irradiated with (a) 0.4 eV/atom, and (b) 0.6 eV/atom absorbed dose, 10 fs FWHM, 30 eV photon energy pulse. Large yellow balls are Cd, small green balls are S.*

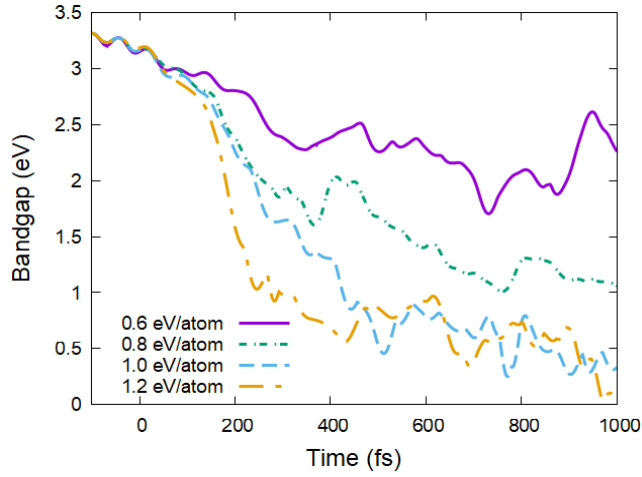

*Figure 7. Band gap of Wurtzite CdS irradiated with various doses.*

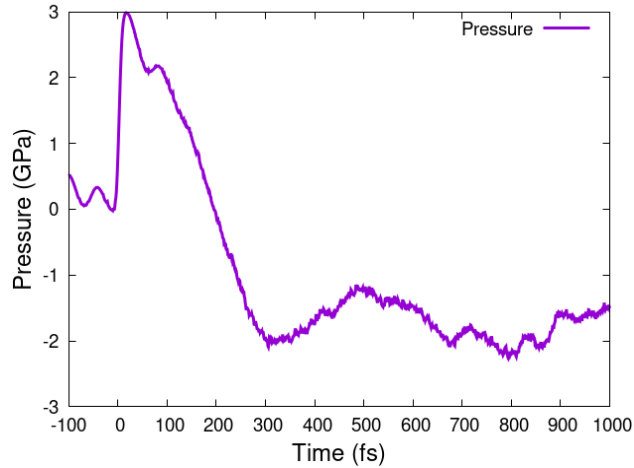

*Figure 8. Pressure in Wurtzite CdS irradiated with 1 eV/atom dose.*

## IV.   Discussion

Thus far, we discussed only the transient states in CdS produced by ultrafast irradiation. Then, the question arises: are those states stable or what will be the final observable phase after material cooling and relaxation? It is known from previous research that depending on the material properties, damage induced by irradiation may become permanent after cooling, or partially or even completely recover [44].

To estimate the possible effect of damage recovery in CdS, we performed a long simulation of 15 ps allowing for cooling *via* Berendsen thermostat with the room temperature of the bath and the characteristic time of 1 ps (1000 fs) [22]. Although this cooling time is rather arbitrary, we note that in experiments, the cooling speed depends on the energy drain from the radiation-affected zone. Thus, it may be controlled by varying the radiation wavelength or incidence angle [45]. Creating short attenuation depth may achieve fast cooling time of the heated region [45].

We simulated the deposited doses up to 10 eV/atom, to ensure that the material damage is reached even though energy is being drained from the system by the thermostat. With the thermostat with 1 ps cooling time, the damage threshold increases to the values around 1.2-1.3 eV/atom (cf. 0.4-0.5 eV/atom in simulations without cooling in the previous section). This indicates the importance of accounting for the heat sinks in experiments if significant temperature gradients are present inducing heat diffusion out of the heat-affected-zone.

In the simulation with the thermostat, the temperatures of zinc blende CdS are decreasing, and the atomic temperature reaches the room temperature by the end of the simulation, see example of 2 eV/atom simulation in Figure 9. It is accompanied by atoms settling in the amorphous phase (see insets in Figure 11 below).

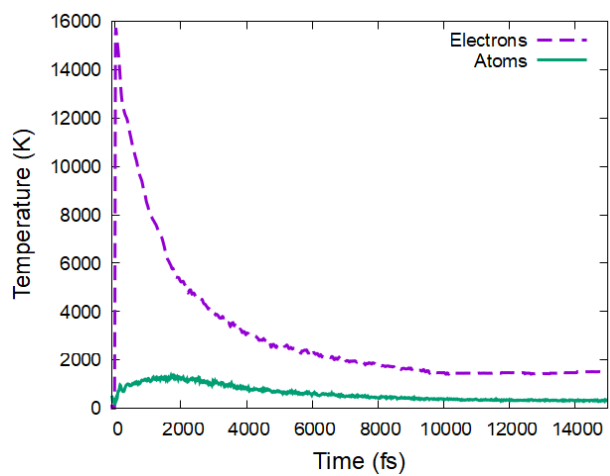

*Figure 9. Electronic and atomic temperatures in zinc blende CdS irradiated with 2 eV/atom dose, cooled down via thermostat with the characteristic time of 1 ps.*

At the doses below ~1.2 eV/atom, the band gap first collapses, as discussed in the previous section, but then opens again with material cooling and recrystallization returning to the original value. In contrast, at higher doses, the band gap settles at smaller values by the time of ~10 ps, see Figure 10. The stable value of the band gap of $E_{gap}$~1.7 eV is created for all the simulated doses above ~1.5 eV/atom, indicating that this is the equilibrium value of the cooled amorphous phase. Interestingly, there is a narrow region of the doses (~1.3 to 1.5 eV/atom) in which the created state has band gap values in between those of the crystalline and stable amorphous phases. This suggests that in experiments the band gap of CdS may be tuned to some degree by tailoring the radiation parameters of the laser pulse.

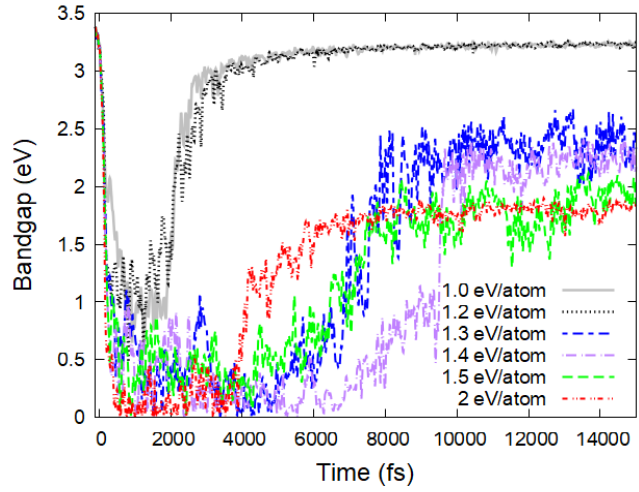

*Figure 10. Band gap values in zinc blende CdS irradiated with various doses, cooled down via thermostat with the characteristic time of 1 ps.*

Produced atomic states may be probed in experiments, e.g., by means of ultrafast X-ray or electron diffraction [46–48]. Thus, we calculated the powder diffraction patterns of CdS irradiated with the dose of 2 eV/atom for the probe photon wavelength of 1.54 Å, see Figure 11. The figure shows that the crystalline peaks start to disappear at the timescale of a few hundred femtoseconds after the pulse arrival, consistent with the timescales of nonthermal melting [14,15]. Then, as the material cools down and relaxes to equilibrium at the timescales of a few picoseconds, some broad diffraction peaks at small angles return, as expected for an amorphous phase. Such noticeable changes in the diffraction patterns should be observable in experiments, thus enabling validation of our predictions.

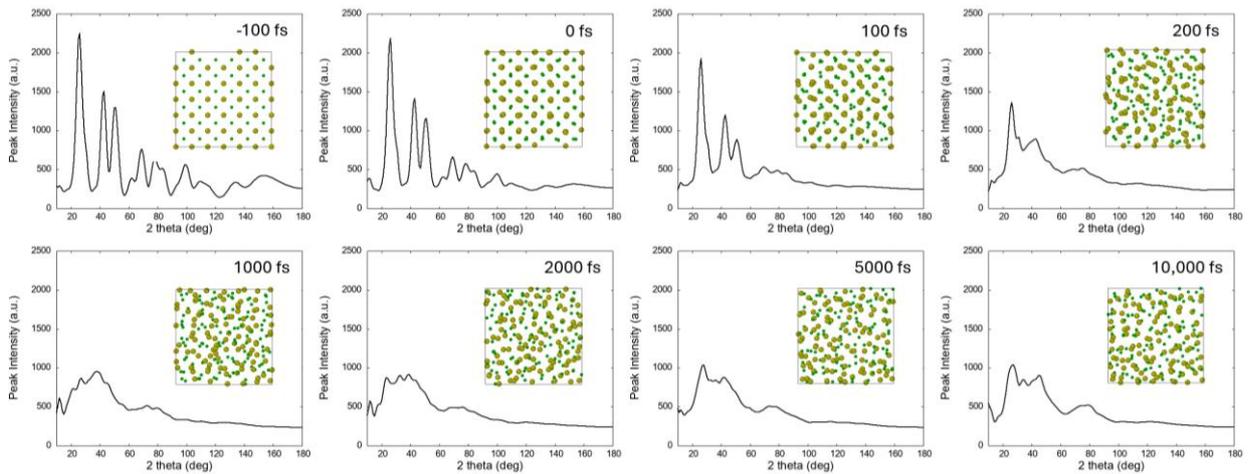

*Figure 11. Powder diffraction patterns (probe wavelength of 1.54 Å) in zinc blende CdS irradiated with 2 eV/atom dose, cooled down with the characteristic time of 1 ps. The insets show the corresponding atomic snapshots.*

# V. Conclusions

Ultrafast XUV/X-ray irradiation of CdS was modelled with the state-of-the-art hybrid code XTANT-3. The two most common phases of CdS were studied: zinc blende and wurtzite. Both phases transiently disorder at irradiation doses above ~0.4-0.5 eV/atom. The melting induced is mainly thermal, triggered by the electron-phonon coupling heating the atomic system. At near-threshold deposited doses, the timescales required for sufficient energy transfer from the electronic system range from 2 to 3 picoseconds. With an increase in the dose, the damage onsets within 1 ps timescale due to nonthermal processes.

In both materials, at the threshold doses the band gap shrinks to the values of ~1 eV, and fully collapses at longer timescales or higher doses of ~1-1.1 eV/atom transiently forming a metallic liquid. At such doses, the damage is mainly nonthermal, triggered by the changes in the atomic potential due to excitation of electrons (when ~2% of valence electrons are promoted to the conduction band). The liquid phases are predicted to be high-density states, denser than either of the crystalline phases studied.

A long simulation with a thermostat (characteristic cooling time of 1 ps) showed an efficient recrystallization of transiently disordered CdS. At the doses below ~1.3 eV/atom, CdS returns to its original state by the time of ~5-10 ps. With the increase of the dose, the cooled state becomes more amorphous, with correspondingly smaller band gap. At the doses above ~1.5 eV/atom, the final state is completely amorphous with the bandgap of ~1.7 eV. The results indicate that adjusting the irradiation dose of a femtosecond laser pulse may allow for tuning the band gap in a CdS-based material.

# VI. Author contributions (CRediT)

N. Medvedev: conceptualization, investigation, formal analysis, methodology, software, supervision, visualization, writing – original draft, writing – review & editing. A. Artímez Peña: conceptualization, investigation, writing – review & editing.

# VII. Conflicts of interest

There are no conflicts to declare.

# VIII. Data and code availability

The code XTANT-3 used to model X-ray irradiation is available from [19].

# IX. Acknowledgments

We thank Dr. Pranab Sarkar for sharing their skf-files with the DFTB parameterization for CdS. The authors gratefully acknowledge financial support from the European Commission Horizon MSCA-SE Project MAMBA [HORIZON-MSCA-SE-2022 GAN 101131245]. Computational resources were provided by the e-INFRA CZ project (ID:90254), supported by the Ministry of Education, Youth and Sports of the Czech Republic. NM thanks the financial support from the Czech Ministry of Education, Youth, and Sports (grant nr. LM2023068).